\documentclass[prb,preprint]{revtex4-1}
              
\usepackage{graphicx}
\usepackage{amsmath}
\usepackage{amsfonts}

\begin{document}

\title{Klein Paradox in the Pilot Wave Interpretation}
\author{John F. Dodaro}
\email{jfd2114@columbia.edu}
\affiliation{Department of Physics, Columbia University, New York, NY 10027}
\date{\today}

\begin{abstract}
The de Broglie-Bohm pilot wave interpretation of quantum mechanics is shown to provide a consistent explanation for a single relativistic particle (more accurately, a single particle \emph{process} since pair production is addressed). This is accomplished by incorporating the Feynman-Stueckelberg interpretation of anti-matter. The lack of conserved probability and existence of negative energy solutions is studied through this interpretation. We discuss the resolution of Klein's original paradox through the pilot wave interpretation, and similarly analyze the potential barrier problem. Simulations are presented for a wave packet incident on potential steps \& barriers of various heights. 
\end{abstract}

\maketitle

\section{Introduction}

The de Broglie-Bohm pilot wave interpretation,\cite{debroglie1, debroglie2, bohm1, bohm2} also known as Bohmian mechanics, offers an alternate interpretation of quantum mechanics from the Copenhagen standpoint. It is deterministic, involving the trajectories of particles in addition to the wavefunction. While the interpretation contains hidden variables, it is in agreement with Bell's theorem due to explicit non-local effects.\cite{bell} For a relativistic particle, the postulates are: (1) the $\psi$-field satisfies the Dirac equation, (2) the particle's trajectory is determined by a first order equation $\dot{x}^\mu \propto J^\mu$ for conserved current $J^\mu$ depending on $\psi$,\cite{bell, bohm1953} and (3) the statistical ensemble of positions obeys the Born rule: $\rho = |\psi|^2$ for probability density $\rho$ at some initial time slice. The universe is not inherently random in this interpretation, but our ignorance of the initial conditions yields a distribution obeying this limitation.\cite{bohm1, bohm2} The Dirac equation is given by:
\begin{equation*} \label{eq:dirac_eom}
\big[ i \hbar \gamma^\mu \partial_\mu - m c \big] \psi = 0
\end{equation*}  
where the $4\times4$ $\gamma$-matrices satisfy $\gamma^\mu \gamma^\nu + \gamma^\nu \gamma^\mu = 2 \eta^{\mu \nu} \mathbb{I}_4$. We use the metric signature $(+,-,-,-)$, and the representation:
\[ \gamma^0 = \left( \begin{array}{cc}
\mathbb{I}_2 & 0 \\
0 & - \mathbb{I}_2 \end{array} \right) \quad \quad \& \quad \quad  \boldsymbol{\gamma} = \left( \begin{array}{cc}
0 &  \boldsymbol{\sigma} \\
-\boldsymbol{\sigma} & 0 \end{array} \right)\] 
for Pauli matrices $\boldsymbol{\sigma}$. The conserved current is given by $J^\mu = c \bar{\psi} \gamma^\mu \psi$ where $\bar{\psi} \equiv \psi^\dagger \gamma^0$. By the second postulate, the guiding equation is:
\begin{equation} \label{eq:dirac_guiding}
\dot{x}^\alpha =  \frac{  c \ J^\alpha}{ \sqrt{J_\mu J^\mu} } 
\end{equation}
where $\dot{x}^\alpha \equiv \frac{d x^\alpha}{d s}$ for affine parameter $s$, and it follows that $\dot{x}_\alpha \dot{x}^\alpha = c^2$. The particle trajectory is then determined given an initial spacetime point $x^\alpha(0)$ and proper boundary conditions on $\psi$. Using $ \gamma^\mu \gamma^\nu = g^{\mu \nu} \mathbb{I}_4 + \frac{1}{2} [ \gamma^\mu , \gamma^\nu] $ with the Dirac (and adjoint) equation, $J^\mu$ can be decomposed as:
\begin{equation*}
\begin{split}
J^\mu = \frac{i \hbar}{2 m} \Big[ \bar{\psi} (\partial^\mu \psi) - (\partial^\mu \bar{\psi}) \psi  \Big] + \frac{i \hbar}{4 m} \partial_\nu \Big( \bar{\psi} [\gamma^\mu , \gamma^\nu] \psi \Big)
\end{split}
\end{equation*} 
which has a similar form to the non-relativistic current. 

In analogy with the non-relativistic case, we can write the equations of motion in second order form;
from the guiding equation (\ref{eq:dirac_guiding}) we have:
\begin{equation*}
\frac{\textbf{v}}{c} = \frac{\dot{x}^i}{\dot{x}^0} = \frac{\bar{\psi} \boldsymbol{\gamma} \psi}{\psi^\dagger \psi}
\end{equation*}
for $i = 1,2,3 $. Defining the matrices $\boldsymbol{\alpha} \equiv \gamma^0 \boldsymbol{\gamma}$ and $\beta \equiv \gamma^0$, the Dirac \& adjoint equations become:
\begin{equation*}
\begin{split}
\dot{\psi} = - c \boldsymbol{\alpha} \cdot \vec{\nabla} \psi + \frac{m c^2}{i \hbar} \beta \psi \quad \quad \& \quad \quad \dot{\bar{\psi}} = - c \vec{\nabla} \bar{\psi} \cdot \boldsymbol{\alpha} - \frac{m c^2}{i \hbar} \bar{\psi} \beta
\end{split}
\end{equation*}
Differentiating with respect to time yields $\textbf{a} = \textbf{a}_L + \textbf{a}_S$, where $\textbf{a}_L$ \& $\textbf{a}_S$ are defined as:
\begin{equation*}
\textbf{a}_L \equiv  \frac{ (\vec{\nabla} \psi^\dagger \cdot  \boldsymbol{\alpha}  ) \left( c \boldsymbol{\alpha}  - \textbf{v}  \right) \psi  -  \psi^\dagger \left( c \boldsymbol{\alpha}  - \textbf{v} \right) ( \boldsymbol{\alpha} \cdot \vec{\nabla} \psi )}{\psi^\dagger \psi} \ \ \ \ \ \ \& \ \ \ \ \ \ \textbf{a}_S \equiv   \frac{ i m c^2}{ \hbar} \left( \frac{ \bar{\psi} \boldsymbol{\alpha} \psi }{\psi^\dagger \psi} \right)
\end{equation*}

The Feynman-Stueckelberg interpretation of anti-matter\cite{stueckelberg, feynman_1949} has been applied to negative energy states to resolve the Klein paradox.\cite{hansen, holstein} Our purpose is to show how it can be incorporated into the pilot wave interpretation for a consistent explanation of a single relativistic particle. We review the Klein paradox and discuss its resolution through this interpretation in Section \ref{section:paradoxA}. Numerical simulations of a Gaussian packet incident on potential steps of various heights are presented in Section \ref{section:paradoxB}. The potential barrier is analyzed in a similar context in Section \ref{section:barrier}, and numerical simulations are presented in Section \ref{section:barrier_sim}.

\section{Klein Paradox} \label{section:paradoxA}

We start by reviewing the Klein paradox\cite{klein, bjorken} in one dimension. The potential is:
\begin{displaymath}
V(x) = \begin{cases}
0, & \text{if } x < 0 \ \ \ \ \text{Region I} \\
V, & \text{if } x > 0 \ \ \ \ \text{Region II}
\end{cases}
\end{displaymath} 
with $V>2mc^2$ and $m<E<V-mc^2 $. Setting $\hbar = c = 1$, defining $\alpha = \gamma^0 \gamma^3$ and $\beta = \gamma^0$, the time independent free Dirac equation for a spin $+\frac{1}{2}$ particle can be written as a two-component spinor:
\begin{equation*}
E \psi = - i \alpha \partial_z \psi + m \beta \psi
\end{equation*}
\begin{eqnarray*}
E \left(
\begin{array}{c}
\phi_+ \\
0 \\ 
\phi_- \\
0 \\
\end{array}
\right) = - i  \alpha \partial_z \left(
\begin{array}{c}
\phi_+ \\
0 \\ 
\phi_- \\
0 \\
\end{array}
\right) + m \beta  \left(
\begin{array}{c}
\phi_+ \\
0 \\ 
\phi_- \\
0 \\
\end{array}
\right)
\end{eqnarray*}
\begin{equation*}
\begin{split}
\Rightarrow \ \ E \left(
\begin{array}{c}
\phi_+ \\
\phi_- \\
\end{array}
\right) = - i  \sigma_x \partial_x \left(
\begin{array}{c}
\phi_+ \\
\phi_- \\
\end{array}
\right) + m \sigma_z  \left(
\begin{array}{c}
\phi_+ \\
\phi_- \\
\end{array}
\right) 
\end{split}
\end{equation*}
For $x<0$, we plug in $\left( \begin{array}{c}
a \\
b \\
\end{array} \right) e^{ + i \tilde{p} x}$ for the incoming and reflected solutions:
\begin{equation*}
\begin{split}
(E-m) a = \tilde{p} b\quad \quad \& \quad \quad (E+m) b = \tilde{p} a
\end{split}
\end{equation*}
such that $\tilde{p} = \pm \sqrt{E^2-m^2}$ and $b=a \frac{\tilde{p}}{ E+m}$. Defining $p \equiv +\sqrt{E^2-m^2}$, the solution in Region I is given by:
\begin{equation*}
\psi_\text{I} = A \left( \begin{array}{c}
1 \\
\frac{p}{E+m} \\
\end{array} \right) e^{ + i p x} + R \left( \begin{array}{c}
1 \\
\frac{-p}{E+m} \\
\end{array} \right) e^{- i p x}
\end{equation*}
For $x>0$, we plug in $\left( \begin{array}{c}
a \\
b \\
\end{array} \right) e^{ + i \tilde{k} x}$ for transmitted wave:
\begin{equation*}
(E-V-m) a =  \tilde{k} b \quad \quad \& \quad \quad (E-V+m) b =  \tilde{k} a
\end{equation*}
such that $\tilde{k} = \pm \sqrt{(E-V)^2-m^2}$. Similarly by defining $k \equiv + \sqrt{(E-V)^2-m^2}$, the solution is:
\begin{equation*}
\psi_\text{II} = T \left( \begin{array}{c}
1 \\
\frac{ k }{ E - V + m  } \\
\end{array} \right) e^{ + i k x } 
\end{equation*}
Continuity at the boundary $\psi_\text{I}(0)=\psi_\text{II}(0)$ yields:
\begin{equation*} 
\kappa \equiv \frac{ k }{ p } \frac{ E + m }{ E - V + m }
\end{equation*}
\begin{equation*}
A + R = T\quad \quad \& \quad \quad A - R = \kappa T  \\
\end{equation*}
We can compute $R$ \& $T$ as:
\begin{equation*}  \label{eq:coefficients_A}
R =  A \left( \frac{ 1 - \kappa }{ 1 + \kappa }  \right)  \quad \quad \& \quad \quad T =  A \left( \frac{ 2 }{ 1 + \kappa }  \right) 
\end{equation*} 
and $|A|^2- \kappa |T|^2 = |R|^2$. The interpretation of probability will be given later. As $V \rightarrow \infty$, $\kappa$ approaches a finite non-zero value (since $k \propto V$) suggesting that an infinite potential has some transparency for the electron. For the current $J^\mu = (\psi^\dagger \psi, \psi^\dagger \sigma_x \psi)$ in Region I:
\begin{equation*}
\psi^\dagger \psi = 1 + \left( \frac{p}{E+m} \right)^2 \quad \quad \& \quad \quad \psi^\dagger \sigma_x \psi = \pm 2 \left( \frac{p}{E+m} \right)
\end{equation*}
\begin{equation*}
\sqrt{J_\mu J^\mu} =   1 - \left( \frac{p}{E+m} \right)^2
\end{equation*}
for incident ($+$) \& reflected ($-$) waves. Using Eq.~(\ref{eq:dirac_guiding}) it follows:
\begin{equation*}
\dot{x}_\text{I}^0 = \frac{d t }{d s} = \frac{ E}{m} \quad \quad \& \quad \quad  \dot{x}_\text{I}^1 = \frac{d x }{d s}  = \pm \frac{p}{m} \quad \quad \Rightarrow \quad \quad \frac{d x}{d t} = \pm \frac{p}{E}  
\end{equation*}
In Region II we have:
\begin{equation*}
\psi^\dagger \psi = 1 + \left( \frac{k}{E-V+m} \right)^2 \quad \quad \& \quad \quad \psi^\dagger \sigma_x \psi =  \frac{ - 2 k }{V-E-m}
\end{equation*}
\begin{equation*}
\sqrt{J_\mu J^\mu} =  1 - \left( \frac{k}{E-V+m} \right)^2
\end{equation*}
for the transmitted wave. Using Eq.~(\ref{eq:dirac_guiding}) we find:
\begin{equation*}
\dot{x}_\text{II}^0 = \frac{d t}{d s}  = \frac{V-E}{m} > 0 \quad \quad \& \quad \quad \dot{x}_\text{II}^1 = \frac{d x }{d s}  = - \frac{k}{m} 
\end{equation*}
The Feynman-Stueckelberg interpretation (of anti-matter as matter going backwards in time) enters when the particle is in Region II since it is a negative energy solution. To keep $E t$ increasing in the phase of $e^{- i E t}$ when $E<0$, we change $+t \rightarrow -t$ (or $+x^0 \rightarrow - x^0$) in such regions. Applying this change to the wavefunction yields $\widetilde{\psi}_{\text{II}}$ satisfying the time-reversed Dirac equation\cite{bjorken}:
\begin{equation*}
\widetilde{\psi}_{\text{II}}(\textbf{x},-t) \equiv \gamma^1 \gamma^3 {\psi}^\dagger_{\text{II}}(\textbf{x},t) \quad \quad \Rightarrow \quad \quad \widetilde{\psi}_\text{II}(x) = T \left( \begin{array}{c}
1 \\
 \frac{  k }{ E - V + m  } \\
\end{array} \right) e^{ - i k x } 
\end{equation*}
which corresponds to the complex conjugate in one dimension; this doesn't affect $\dot{x}_\text{II}^0$, $\dot{x}_\text{II}^1$, or the correct boundary condition $\psi_\text{I}(0)=\widetilde{\psi}_\text{II}(0)$ since the phase cancels out. Applying the change $+t \rightarrow - t$ to the guiding equation (\ref{eq:dirac_guiding}) in Region II yields:
\begin{equation*}
\dot{x}_\text{II}^0 = \frac{d t}{d s}  = \frac{E-V}{m} < 0 \quad \quad \& \quad \quad \dot{x}_\text{II}^1 = \frac{d x }{d s}  = - \frac{k}{m} \quad \quad \Rightarrow \quad \quad \frac{d x}{d t} = \frac{k}{V-E} > 0 
\end{equation*}
This changes the sign of $d t /d s$, and therefore $d x / d t$. The direction of the velocity $d x / d t$ in the lab frame is to the \emph{right} even though the particle's velocity (with  monotonically increasing affine parameter $s$) is to the \emph{left} and backwards in time. The process can be described as packet $A$ at $t=-\infty$ \& $x=-\infty$ and packet $T$ at $t=+\infty$ \& $x=+\infty$. Packet $A$ moves forward in time, packet $T$ moves backwards in time, and both meet at the step boundary. They constructively interfere forming the reflected packet $R$ which moves forward in time from the step boundary to $x=-\infty$.  

Applying the pilot wave interpretation, there is a probability that the particle is in either packet $A$ or packet $T$ (even though the two packets aren't on the same time slice) with a definite, continuous trajectory \emph{over all} of spacetime. The particle's initial spacetime position used to uniquely solve the guiding equation is defined at $s=-\infty$; since $d t / d s > 0$ in Region I, $t \rightarrow -\infty$ as $s \rightarrow - \infty$ implying an ensemble of initial positions in packet $A$. In Region II, $d t / d s < 0$ such that $t \rightarrow +\infty$ as $s \rightarrow - \infty$; this implies the ensemble also includes a distribution of initial positions in packet $T$ which will travel backwards in time by the guidance condition. 

If the particle starts in packet $A$, it is reflected at the step boundary; if the particle starts in packet $T$, it goes backwards in time to the step boundary, and then moves forward in time with packet $R$ towards $x=-\infty$. In the lab frame, this would appear to describe the creation of a particle/anti-particle pair at the step boundary. As there are two particles in the lab frame, it would be more accurate to describe the ensemble as a single particle \emph{process} since there is only a single, continuous trajectory over all of spacetime. If the particle's initial spacetime point corresponds to reflection, the overall charge is $-1$ (for electrons); if the initial spacetime point corresponds to a trajectory backwards in time, the overall charge is $0$ since in the lab we observe pair creation at the step boundary. It follows that charge is conserved on each time slice.

Returning to the interpretation of probability, we see at $s=-\infty$, there are two spacetime packets containing the ensemble with coefficients $A$ \& $T$. Solving for $A$ \& $T$:
\begin{equation*}
A = R \left( \frac{1 + \kappa}{1 - \kappa}  \right) \quad \quad \& \quad \quad T =  R \left( \frac{ 2 }{1 - \kappa} \right) 
\end{equation*}
As $s \rightarrow \infty$, all the particles move to the packet with coefficient $R$. Since $\kappa < 0$, the relation:
\begin{equation*}
|A|^2- \kappa |T|^2 = |R|^2
\end{equation*} 
implies the conservation of \emph{spacetime} probability with respect to $s$; $ |A|^2 / |R|^2$ is probability of the particle starting in packet $A$ at $s=-\infty$, and $ -\kappa |T|^2 / |R|^2 $ is the probability of the particle starting in packet $T$ at $s=-\infty$. It follows that the average number of particle/anti-particle pairs produced at the step boundary is given by\cite{footnote1}:
\begin{equation*} 
\left( \frac{ - \kappa |T|^2}{|R|^2} \right) |R|^2 = \frac{ - 4 \kappa }{(1-\kappa)^2} |R|^2 =   \frac{ - 4 \kappa }{(1+\kappa)^2} |A|^2 
\end{equation*}
We can now reinterpret Klein's result: as $V \rightarrow \infty$, $- \kappa |T|^2$ approaches a finite non-zero value. This implies that for a potential $ V > 2 m c^2 $, there will be pair production at the step boundary even if $V \rightarrow \infty$. The new feature of the interpretation is the probability of an \emph{initial} particle position at a \emph{future} time slice. As we shall see in Section \ref{section:barrier}, this isn't encountered in the potential barrier case. The feature is due to the infinite extent of the potential $V > 2 m c^2 $ to $x=+\infty$ yielding an asymptotic negative energy solution.

\section{Numerical Analysis - Step} \label{section:paradoxB}

We present numerical simulations for an incident Gaussian packet starting with the free particle (Case 0). The three cases with $V \neq 0$ are: $V+mc^2 < E$ (Case 1), $V-mc^2<E<V+mc^2$ (Case 2), and the Klein paradox $mc^2<E<V-mc^2$ (Case 3). Setting $\hbar=c=m=1$, the packet is constructed by integrating plane wave solutions over energy. Following the notation of Nitta, Kudo, and Minowa,\cite{nitta} we start by defining:
\begin{equation*}
\phi_E(x) = \psi_{\text{I}}(x) \Theta(-x) + \psi_{\text{II}}(x) \Theta(x)
\end{equation*}
where $\Theta(x)$ is the Heaviside step function. The forward evolving solution $\psi_{\text{II}}$ (and corresponding boundary condition) is replaced by the time-reversed wavefunction $\widetilde{\psi}_{\text{II}}$ for Case 3. The time-dependent wavefunction $\Psi(x,t)$ is:
\begin{equation*}
\Psi(x,t) = \int_D G \left( \sqrt{E^2-m^2} \right) \phi_E(x) e^{- i E t} dE
\end{equation*}
where $D_1 = \{V+m<E<V+2m\}$ for Case 1, $D_2 = \{V-m<E<V+m\}$ for Case 2, and $D_3 = \{ m < E < V-m \}$ for Case 3. The Gaussian function $G(p)$ is defined as:
\begin{equation*}
G(p) = e^{- \frac{\lambda^2}{2} (p-K_0)^2 - i p X_0 }
\end{equation*}
for mean momentum $K_0$, mean position $X_0$, and position spread $\lambda$. The mean momentum is taken to be $K_0  = 1 / \sqrt{3} $ (except for Case 0) such that the mean incident velocity is $v = 1 / 2 $, and the mean energy is $E = 2 / \sqrt{3}$.

\begin{figure}[h!] 
\centering
\includegraphics [width=0.6 \textwidth]{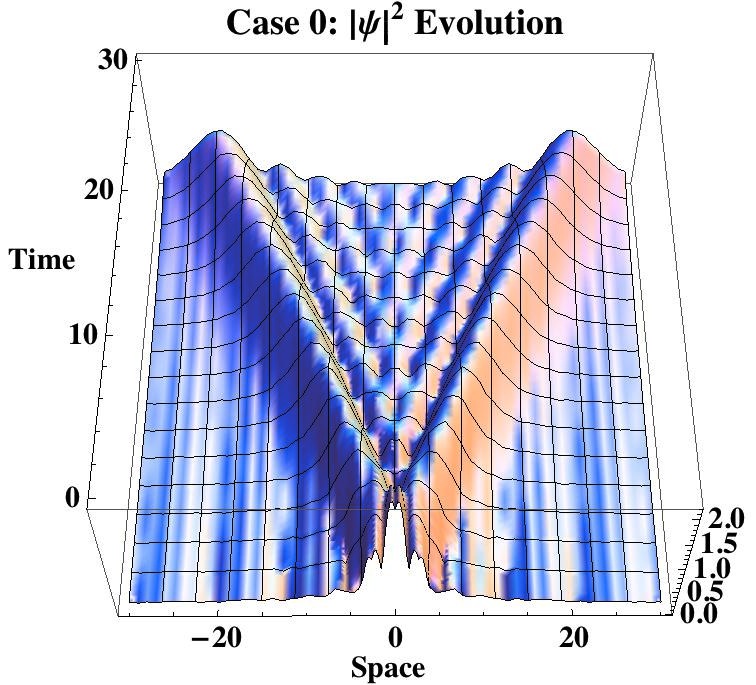}
\caption{(Case 0) Probability evolution of packet with $V=0$.}
\label{fig:free1}
\end{figure}

\begin{figure}[h!]
\centering
\includegraphics[width=0.55 \textwidth]{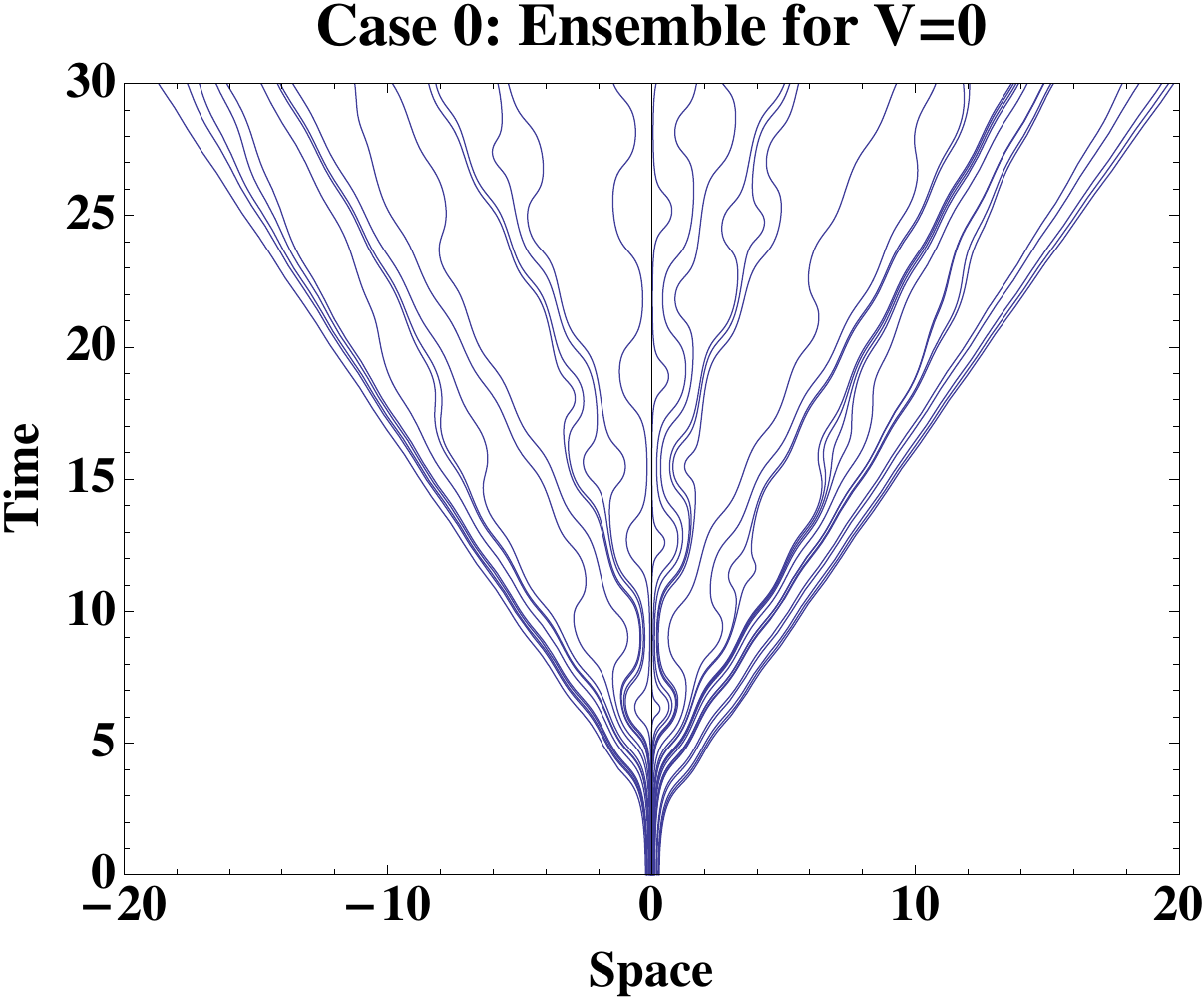}
\caption{(Case 0) Ensemble of trajectories for $V=0$ with initial conditions at $t=0$ for random Gaussian distribution. The \emph{zitterbewegung} motion is most apparent for particles closer to rest (center), and is barely visible for particles closer to the speed of light (edges).}
\label{fig:free2}
\end{figure}

We first consider the case of $V=0$, $K_0=0$, $\lambda=0.1$, $T=0$, and $A=R=1$. The probability density is plotted (Figure \ref{fig:free1}), and the trajectories for a random Gaussian distribution are computed (Figure \ref{fig:free2}). Since $\lambda < \hbar/ m c = 1$ by our choice of units, the initial wavefunction is confined to a width smaller than the Compton wavelength causing two packets to move out near the speed of light. The \emph{zitterbewegung} motion visible in the trajectories is due to the interference between positive and negative energy states. In general the trajectories may converge, but do not cross; this is expected by uniqueness from the first order guiding equation in one spatial dimension.

\begin{figure}[h!]
\centering
\includegraphics[width=0.65 \textwidth]{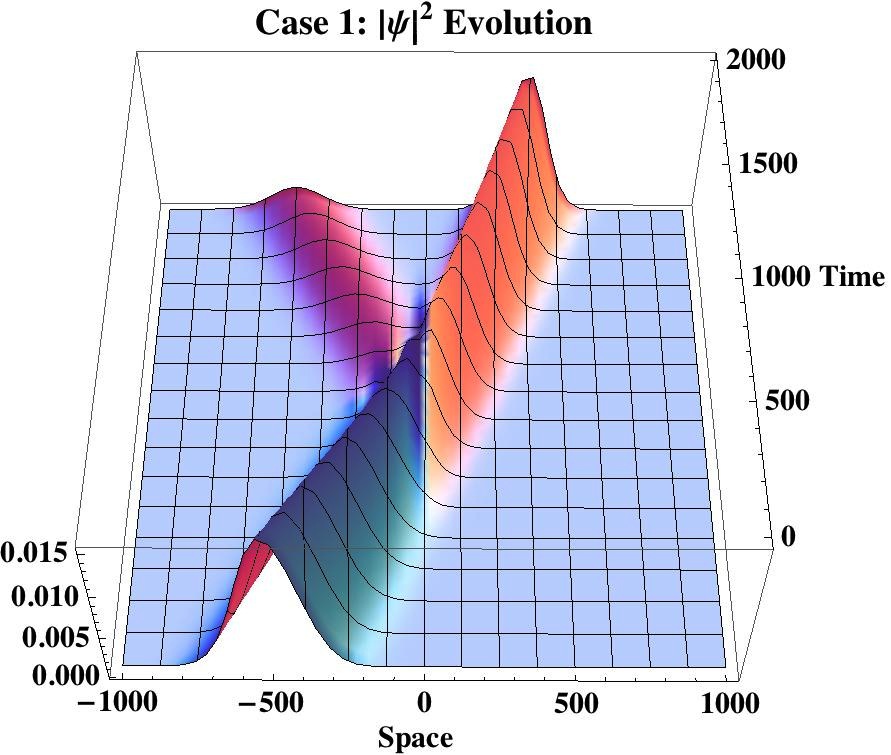}
\caption{(Case 1) Probability evolution of packet incident on step with $V+m<E$.}
\label{fig:step1}
\end{figure}

\begin{figure}[h!] 
\centering
\includegraphics[width=0.5 \textwidth]{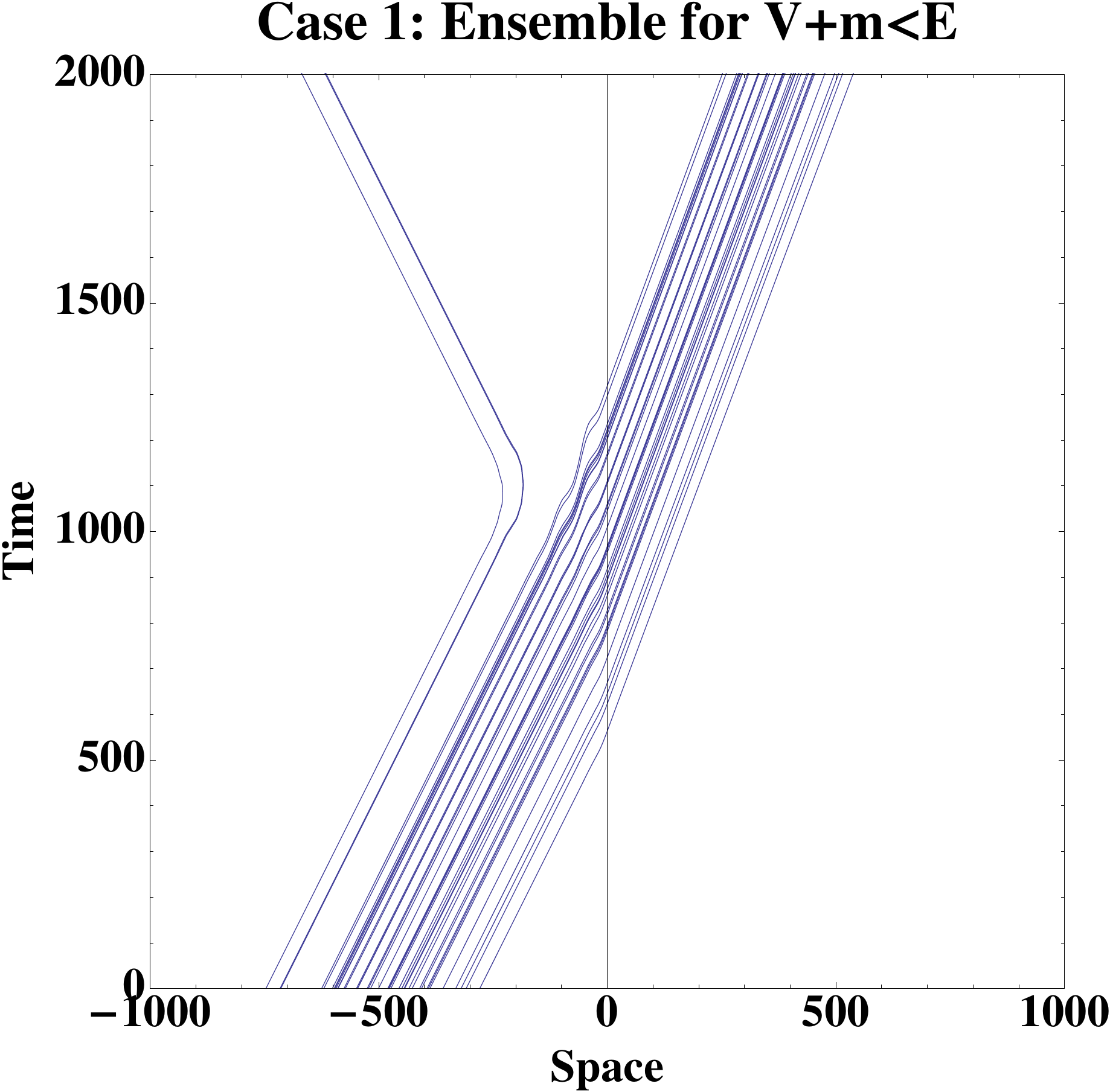}
\caption{(Case 1) Ensemble of trajectories for $V+m<E$ with initial conditions at $t=0$ for random Gaussian distribution.}
\label{fig:step2}
\end{figure}

Next we consider Case 1: $V+m < E$ (Figure \ref{fig:step1} \& \ref{fig:step2}) taking $V=(E-m)/2=1/\sqrt{3}-1/2$ and $\lambda=100$. Similar to non-relativistic tunneling,\cite{norsen} particles starting closer to the step continue through it, and particles starting further from the step may never reach it (and are reflected even though $V+m<E$). This must be the case, again by uniqueness. There is a clear bifurcation point between reflected \& transmitted trajectories based on initial position. 

\begin{figure}[h!]
\centering
\includegraphics[width=0.65 \textwidth]{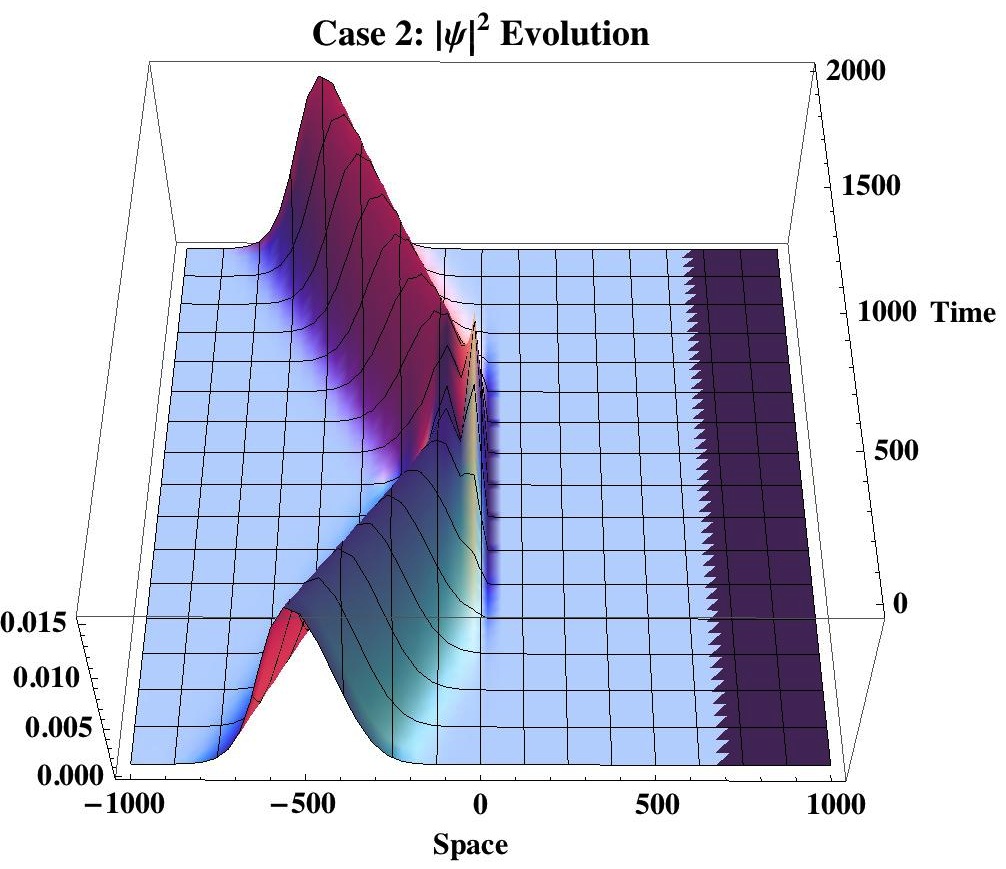}
\caption{(Case 2) Probability evolution of packet incident on step with $V-m<E<V+m$.}
\label{fig:decay1}
\end{figure}

\begin{figure}[h!] 
\centering
\includegraphics[width=0.5 \textwidth]{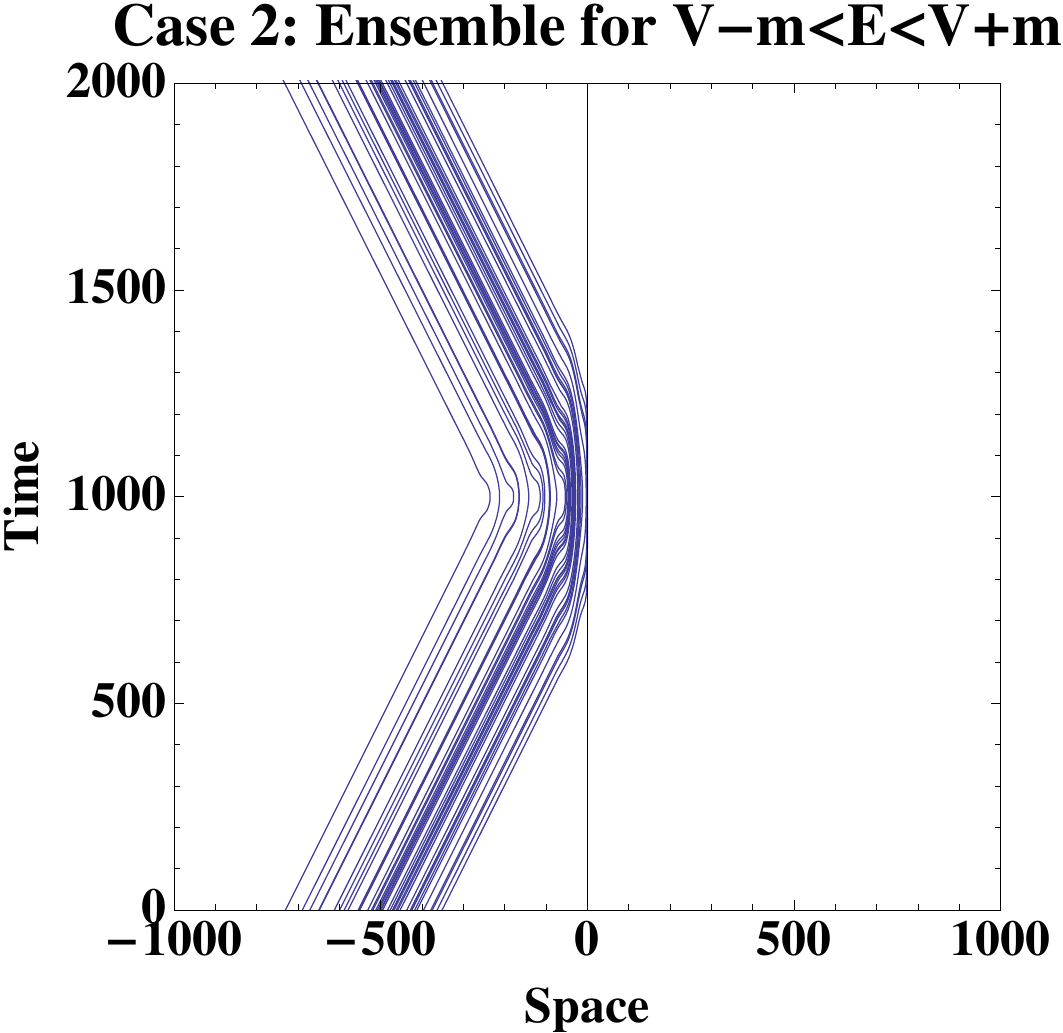}
\caption{(Case 2) Ensemble of trajectories for $V-m<E<V+m$ with initial conditions at $t=0$ for random Gaussian distribution.}
\label{fig:decay2}
\end{figure}

We now consider Case 2: $V-m < E < V+m$ (Figure \ref{fig:decay1} \& \ref{fig:decay2}) with parameters $V=2$ and $\lambda=100$. The potential is chosen such that the negative energy continuum is not opened up. The solution corresponds to an exponential decay for $x>0$, and therefore total reflection which can be seen in the trajectories.

\begin{figure}[h!] 
\centering
\includegraphics[width=0.65 \textwidth]{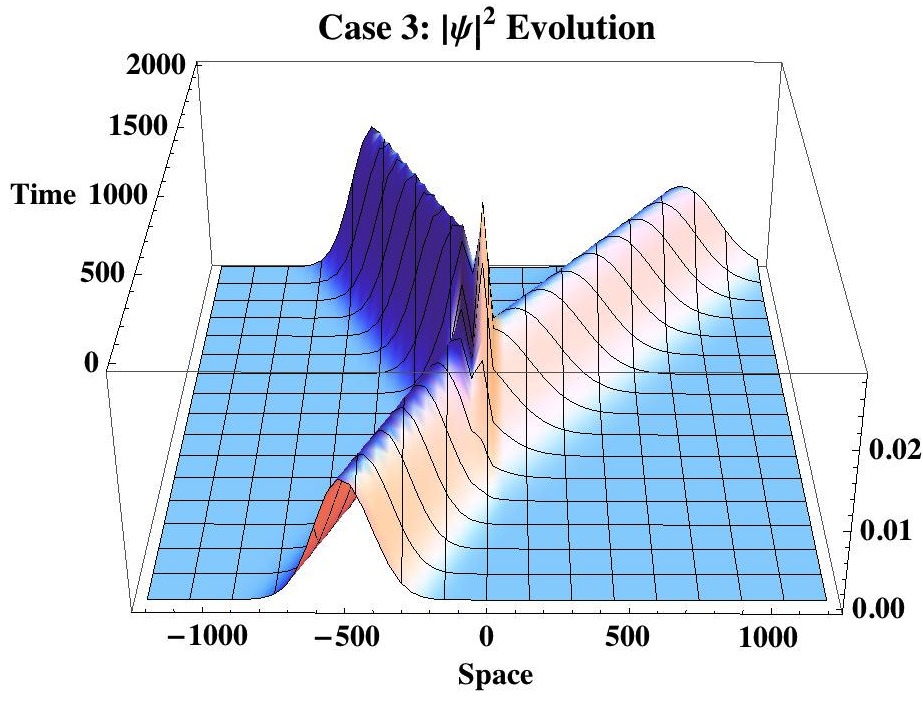}
\caption{(Case 3) Probability evolution of packet incident on step with $m<E<V-m$.}
\label{fig:klein1}
\end{figure}

\begin{figure}[h!] 
\centering
\includegraphics[width=0.55 \textwidth]{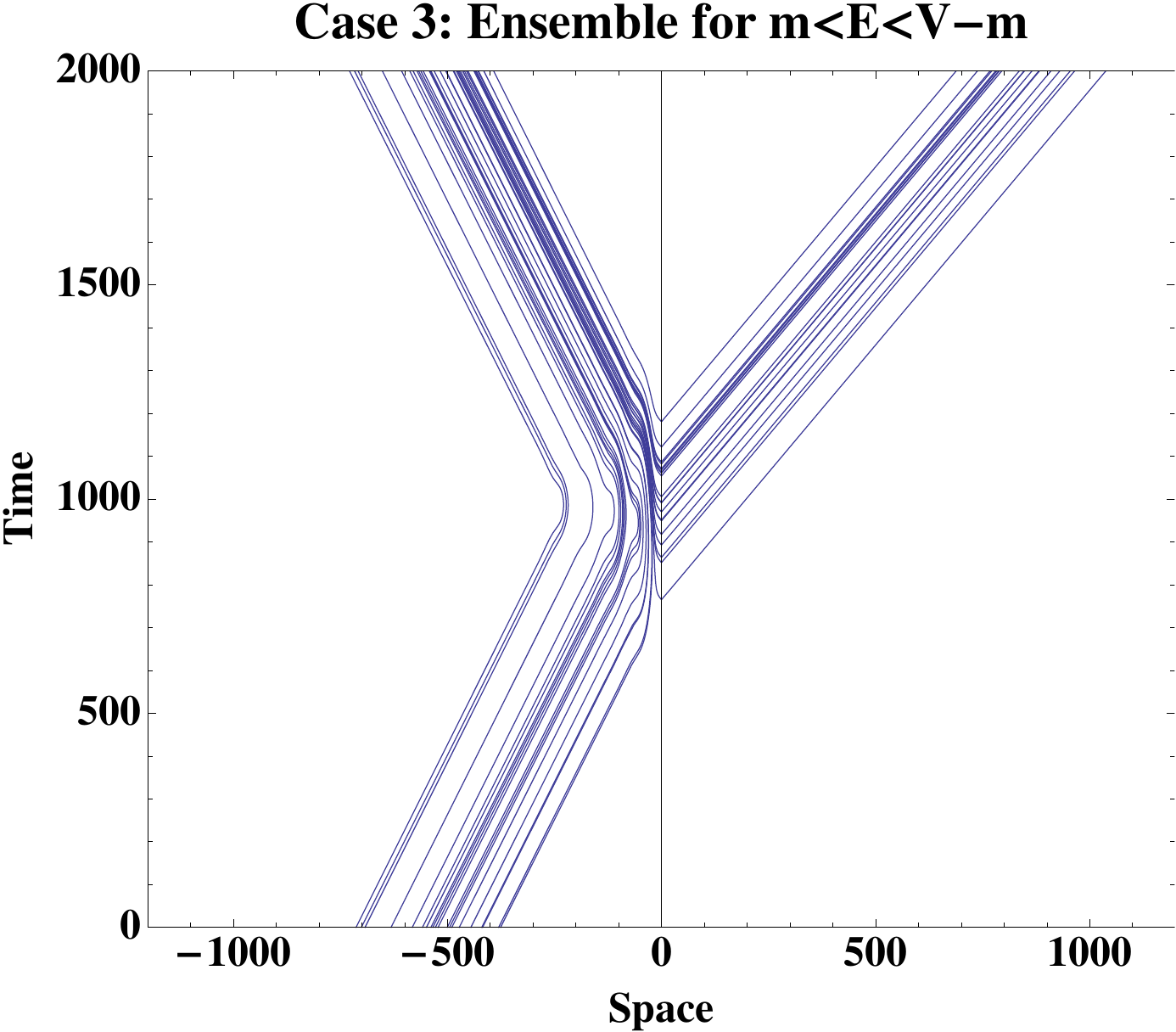}
\caption{(Case 3) Ensemble of trajectories for $m<E<V-m$ with initial conditions at $t=0$ \& $t=\tau_F$ for random Gaussian distribution.}
\label{fig:klein2}
\end{figure}

Lastly we consider Case 3: the Klein paradox with $m < E < V-m $ (Figure \ref{fig:klein1} \& \ref{fig:klein2}) for parameters $V=3$ and $\lambda=100$. It can be seen from the probability distribution (Figure \ref{fig:klein1}) that the reflected packet is larger than the incident packet. The probability distributions can be integrated at initial time $t=0$, and final time $t=\tau_F$:
\begin{equation*}
P_A = \int_{-\infty}^0  \Psi^\dagger(x,0) \Psi(x,0)  dx \ ; \quad \quad P_T =  \int_{0}^{+\infty}  \Psi^\dagger(x,\tau_F) \Psi(x,\tau_F)  dx 
\end{equation*}
\begin{equation*}
P_R =  \int_{-\infty}^0  \Psi^\dagger(x,\tau_F) \Psi(x,\tau_F)  dx
\end{equation*}
and the probability conservation $P_A+P_T=P_R$ with respect to affine parameter $s$ can be verified numerically. The initial conditions for the particle distribution are on the $t=0$ and $t=\tau_F$ time slices, and correspond to a random Gaussian centered on the packets. It can be seen from the trajectories (Figure \ref{fig:klein2}) that all incident particles are reflected, and all particles starting at $t=\tau_F$ form a ``\emph{V}'' shape corresponding to pair creation at the boundary. By the choice of mean energy and barrier height, the anti-particles in Region II move at a greater velocity than the particles in Region I.

\section{Potential Barrier} \label{section:barrier}

Now we apply the interpretation to the potential barrier.\cite{thomson} The potential is given by:
\begin{displaymath}
V(x) = \begin{cases}
0, & \text{if } x < 0 \ \ \ \ \ \ \ \ \ \ \text{Region I} \\
V, & \text{if } 0 < x < L \ \ \ \ \text{Region II} \\
0, & \text{if } x > L \ \ \ \ \ \ \ \ \ \text{Region III} 
\end{cases}
\end{displaymath} 
with $V>2m$ and $m<E<V-m $. Defining $p \equiv +\sqrt{E^2-m^2}$, the solution in Region I is: 
\begin{equation*}
\psi_\text{I} = A \left( \begin{array}{c}
1 \\
\frac{p}{E+m} \\
\end{array} \right) e^{ + i p x} + R \left( \begin{array}{c}
1 \\
\frac{-p}{E+m} \\
\end{array} \right) e^{- i p x}
\end{equation*}
Defining $k \equiv + \sqrt{(E-V)^2-m^2}$, in Region II:
\begin{equation*}
\psi_\text{II} = B \left( \begin{array}{c}
1 \\
\frac{ k }{ E - V + m  } \\
\end{array} \right) e^{ + i k x } + D \left( \begin{array}{c}
1 \\
\frac{ - k }{ E - V + m  } \\
\end{array} \right) e^{ - i k x } 
\end{equation*}
\begin{equation*}
\Rightarrow \quad \widetilde{\psi}_\text{II} = B \left( \begin{array}{c}
1 \\
\frac{ k }{ E - V + m  } \\
\end{array} \right) e^{ - i k x } + D \left( \begin{array}{c}
1 \\
\frac{ - k }{ E - V + m  } \\
\end{array} \right) e^{ + i k x } 
\end{equation*}
where $\widetilde{\psi}_\text{II}$ corresponds to the time-reversed solution in Region II. Continuity at the boundary $\psi_\text{I}(0)=\widetilde{\psi}_\text{II}(0)$ yields:
\begin{equation*} 
\kappa \equiv \frac{ k }{ p } \frac{ E + m }{ E - V + m }
\end{equation*}
\begin{equation*}
A  + R  = B + D \quad \quad \& \quad \quad A  - R  =  \kappa \left( B  - D \right)
\end{equation*}
The solution in Region III is given by:
\begin{equation*}
\psi_\text{III} = T \left( \begin{array}{c}
1 \\
\frac{p}{E+m} \\
\end{array} \right) e^{ + i p x} 
\end{equation*}
with boundary condition $\widetilde{\psi}_\text{II}(L)=\psi_\text{III}(L)$ such that:
\begin{equation*}
B e^{- i k L} + D e^{ + i k L} = T e ^{+ i p L} \quad \quad \& \quad \quad \kappa \left( B e^{- i k L} - D e^{+ i k L} \right) = T e^{+ i p L}
\end{equation*}
We can then compute $R$ \& $T$ as:
\begin{equation*}
\frac{R}{A} =  \frac{ \left( 1 - \kappa^2  \right) \left( e^{+ikL} - e^{-ikL} \right) }{ e^{+ikL} (1+\kappa)^2  - e^{-ikL} (1-\kappa)^2  } \quad \quad \& \quad \quad \frac{T}{A} =  \frac{ 4 \kappa e^{- i p L} }{ e^{+i k L}(1+\kappa)^2 - e^{- i k L} (1-\kappa)^2 }
\end{equation*}
and the reflection \& transmission probabilities:
\begin{equation*}
\begin{split}
\frac{|R|^2}{|A|^2} =  \frac{   4 \  \text{sin}^2[ k L]  (1-\kappa^2)^2  }{ \left( 1+\kappa \right)^4+\left( 1-\kappa \right)^4 - 2 \ \text{cos}[2 k L] (1-\kappa^2)^2   } \\
\frac{|T|^2}{|A|^2} = \frac{16 \kappa^2}{(1+\kappa)^4+(1-\kappa)^4- 2 \ \text{cos}[2 k L] (1-\kappa^2)^2 }
\end{split}
\end{equation*}
from which it follows $|R|^2+|T|^2=|A|^2$. Using Eq.~(\ref{eq:dirac_guiding}), the velocity in Region I is:
\begin{equation*}
\dot{x}_\text{I}^0  = \frac{ E}{m} \quad \quad \& \quad \quad \dot{x}_\text{I}^1 = \pm \frac{p}{m} \quad \Rightarrow \quad \frac{d x}{d t} = \pm \frac{p}{E}  
\end{equation*}
Similarly in Region III we have:
\begin{equation*}
\dot{x}_\text{III}^0 =  \frac{ E}{m} \quad \quad \& \quad \quad \dot{x}_\text{III}^1 = + \frac{p}{m} \quad \Rightarrow \quad \frac{d x}{d t} = + \frac{p}{E}  
\end{equation*}
where the solutions correspond to positive energy states in Regions I \& III. In Region II:
\begin{equation*}
\dot{x}_\text{II}^0 = \frac{E-V}{m} < 0 \quad \quad \& \quad \quad \dot{x}_\text{II}^1  = \mp \frac{k}{m} \quad  \Rightarrow  \quad \frac{d x}{d t} =  \frac{ \pm k}{V-E}  
\end{equation*}
after changing $ + t \rightarrow - t$ for the negative energy state. The top signs corresponds to $B$, and bottom to $D$. In Regions I \& III, $t \rightarrow - \infty$ as $s \rightarrow - \infty$ such that the distribution of initial position is at $t=-\infty$ \& $x=-\infty$. In Region II, $t \rightarrow + \infty$ as $s \rightarrow - \infty$; since there is no wave at $t=+\infty$ in Region II,\cite{footnote2} there isn't a distribution of initial positions at $t=+\infty$ as we saw in the Klein step. This implies the entire distribution is in Region I at $t=-\infty$ \& $x=-\infty$. We don't encounter the feature of the Klein paradox (with an ``initial'' distribution of particles at $t=+\infty$) regarding probability since $|R|^2+|T|^2=|A|^2$, but there is still finite transmission as $V \rightarrow \infty$. This again suggests that there will be pair creation at the right side of the barrier even if $V \rightarrow \infty$. 

As pointed out by Thomson and McKellar,\cite{thomson} there is an analogy between the Fabry-Perot interferometer, and the incident wavefunction that enters Region II and moves backwards in time. The packet in Region II will diminish as smaller packets tunnel out of the barrier and move forward in time (the limit of multiple internal reflections is studied in the Appendix). This results in a Fabry-Perot pattern in Regions I \& III at $ t = + \infty$. As we will see in Section \ref{section:barrier_sim}, this pattern is visible in the particle trajectories. 

A comment should be made on the choice of words used to describe this process. The wave packet ``traveling backwards in time'' is used as an interpretational description similar to a particle going ``forwards/backwards'' in time. Like the particle worldline, the wavefunction exists over all of spacetime, and therefore shouldn't be thought of as \emph{dynamically} evolving in spacetime. It is equivalent to think of the wavefunction evolving forward in time from the initial conditions set at $ t = - \infty $, but via the time reversed solution in Region II. The \emph{initial} condition in Region II at $t = - \infty$ can therefore be viewed as a \emph{final} condition being evolved backwards to $t =+ \infty$.

\section{Numerical Analysis - Barrier}  \label{section:barrier_sim}

\begin{figure}[h!]
\centering
\includegraphics[width=0.7 \textwidth]{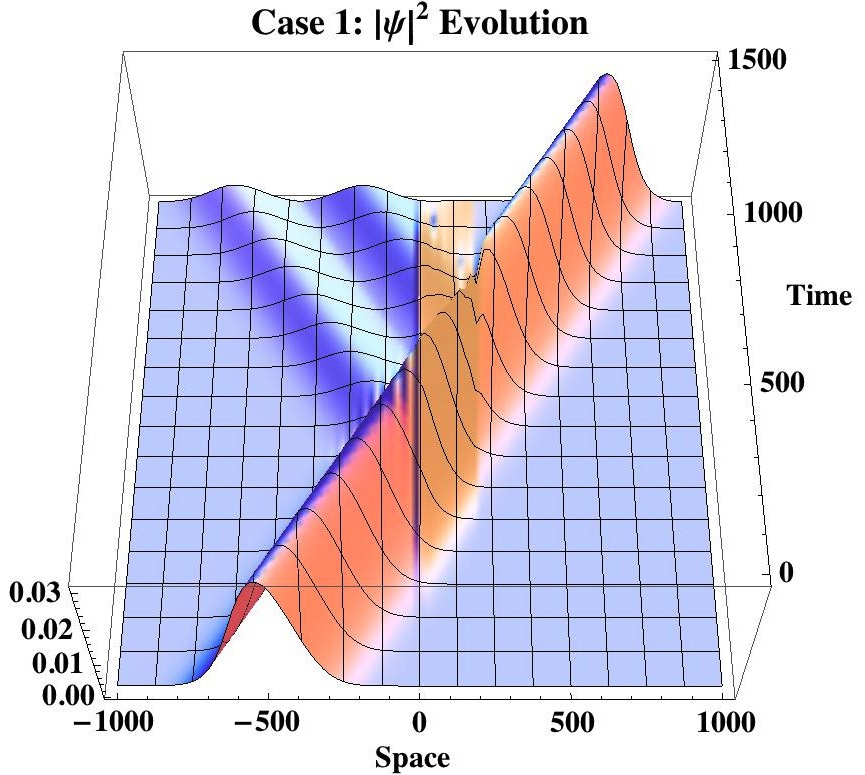}
\caption{(Case 1) Probability evolution of packet incident on barrier with $V+m<E$.}
\label{fig:barrier_step1}
\end{figure}

\begin{figure}[h!] 
\centering
\includegraphics[width=0.6 \textwidth]{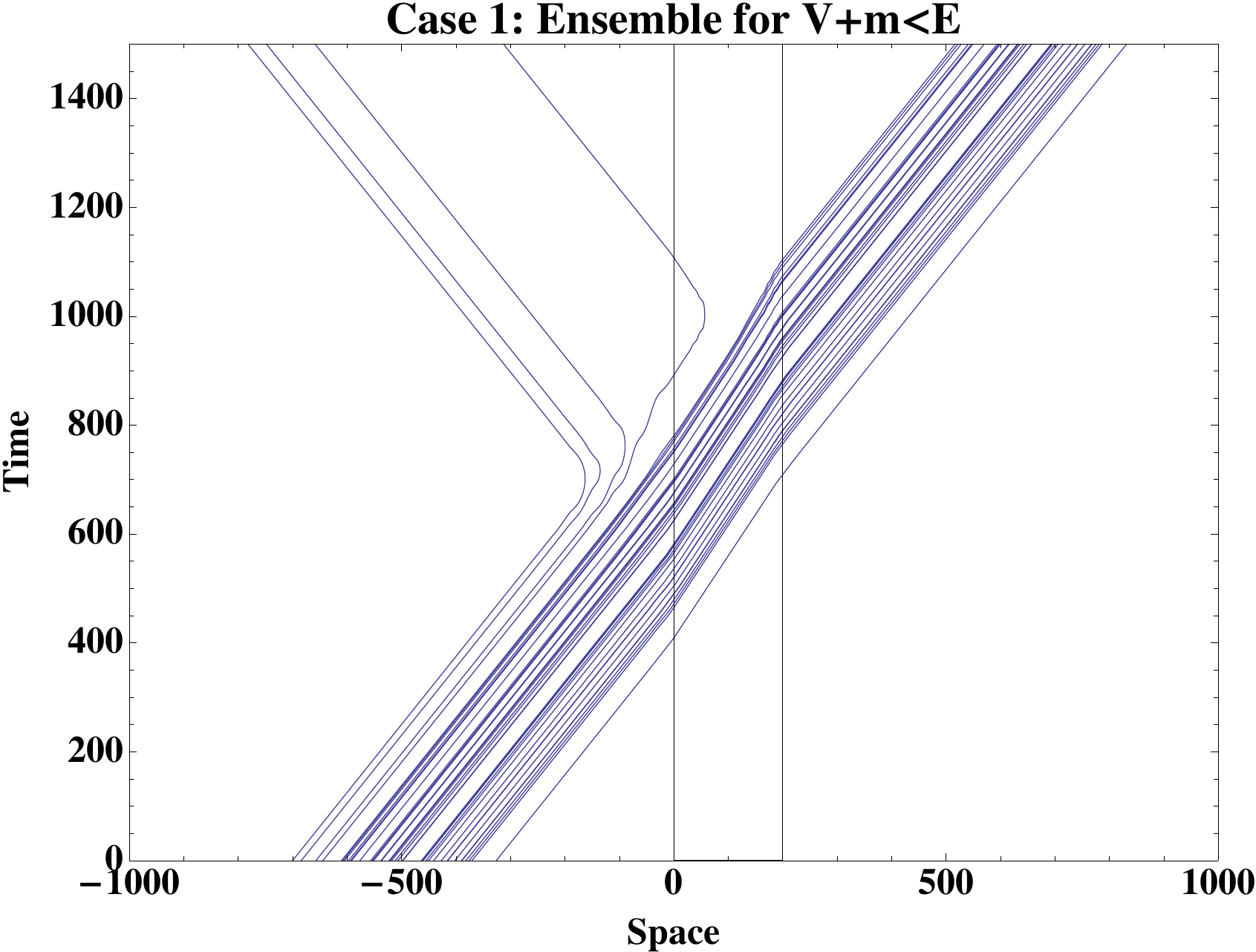}
\caption{(Case 1) Ensemble of trajectories for $V+m<E$ barrier with initial conditions at $t=0$ for random Gaussian distribution.}
\label{fig:barrier_step2}
\end{figure}

A similar numerical analysis is performed on the potential barrier (Cases 1-3) taking $\lambda=100$. We choose $K_0=4/3$ such that the mean incident velocity is $v = 4 / 5 $, and the mean energy is $E = 5 / 3$. The probability distributions are integrated, and the conservation equation $P_R+P_T=P_A$ is verified numerically. We start with Case 1: $V+m<E$ (Figure \ref{fig:barrier_step1} \& \ref{fig:barrier_step2}) for $V=(E-m)/2=1/3$ and barrier width $L=200$. Two reflected packets are visible from scattering off both sides of the barrier. The initial conditions on the $t=0$ time slice correspond to a random Gaussian distribution centered around the packet. The particles slow down inside of the barrier, and reflection still occurs with small probability even though $V+m<E$.

\begin{figure}[h!]   
\centering
\includegraphics[width=0.6 \textwidth]{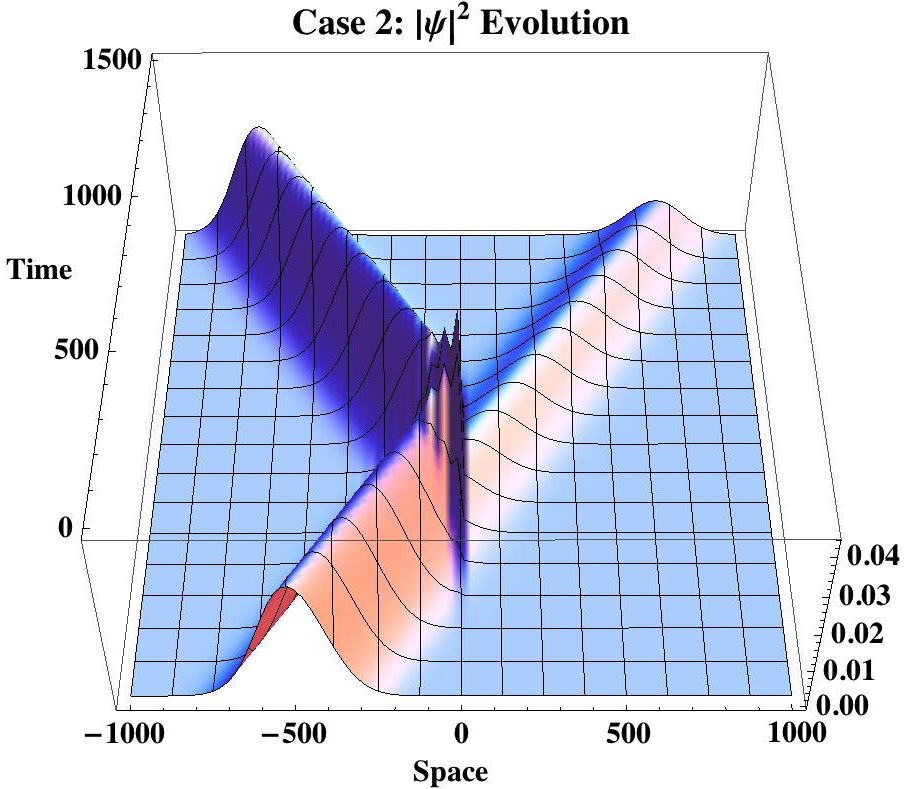}
\caption{(Case 2) Probability evolution of packet incident on barrier with $V-m<E<V+m$.}
\label{fig:barrier_decay1}
\end{figure}

\begin{figure}[h!]   
\centering
\includegraphics[width=0.6 \textwidth]{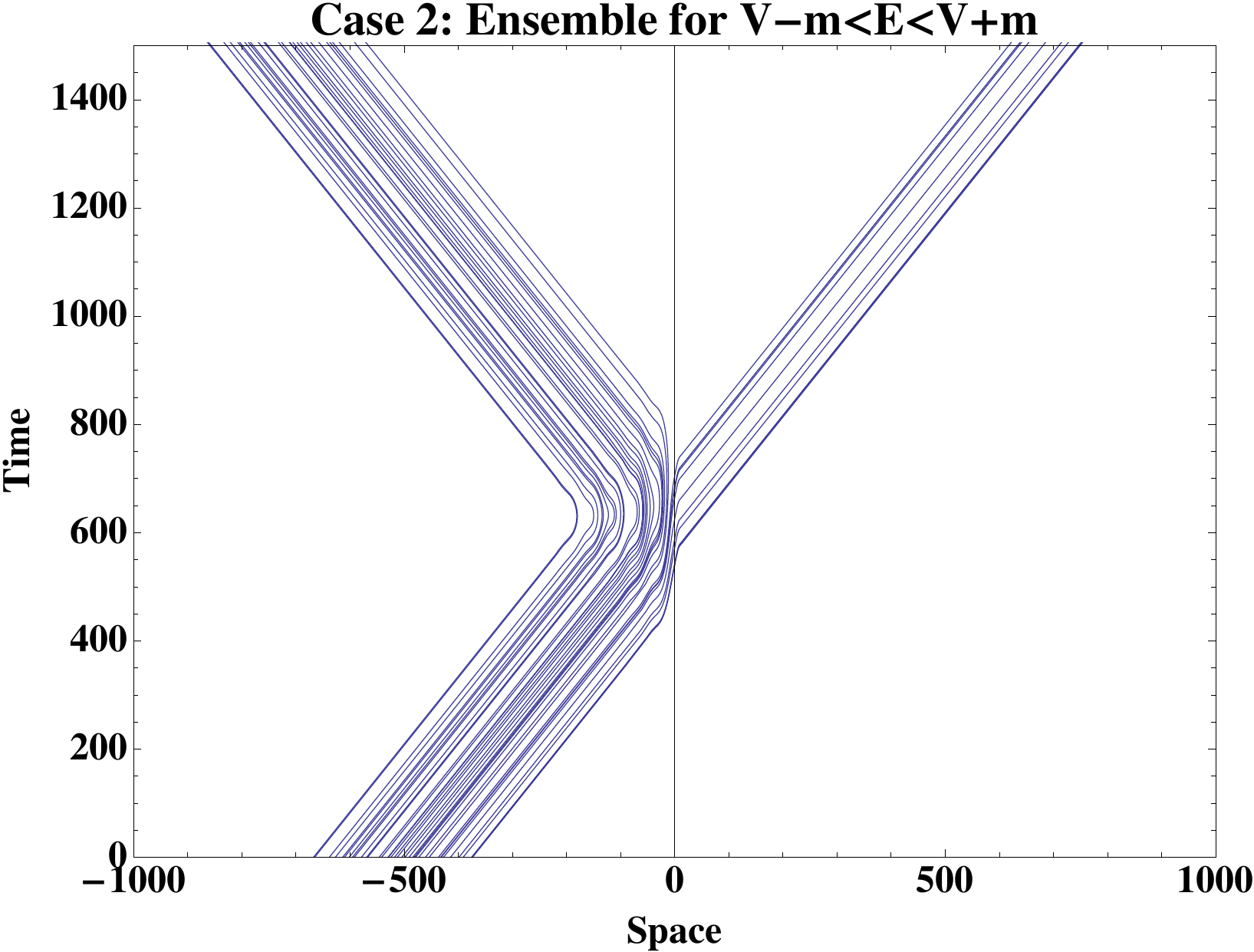}
\caption{(Case 2) Ensemble of trajectories for $V-m<E<V+m$ barrier with initial conditions at $t=0$ for random Gaussian distribution.}
\label{fig:barrier_decay2}
\end{figure}

Next we consider Case 2: $V-m<E<V+m$ (Figure \ref{fig:barrier_decay1} \& \ref{fig:barrier_decay2}) with $V=2$ and barrier width $L=1$. The small width is chose such that a non-negligible packet tunnels through the barrier. This most closely resembles non-relativistic tunneling through a potential barrier\cite{norsen, hiley} since the solution is exponentially decaying inside Region II. 

\begin{figure}[h!] 
\centering
\includegraphics[width=0.65 \textwidth]{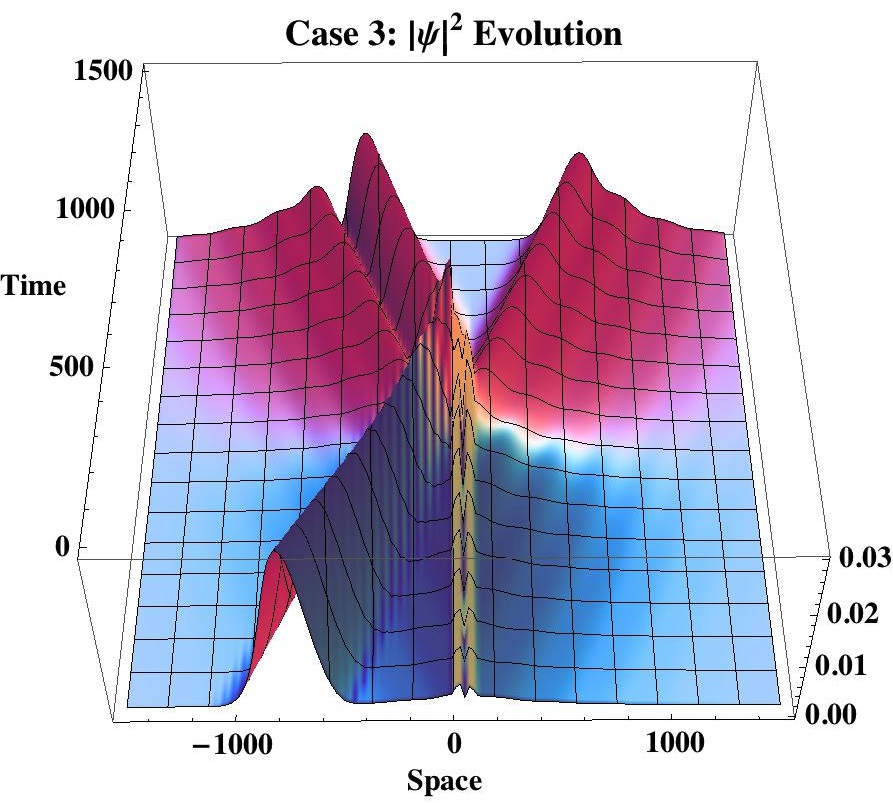}
\caption{(Case 3) Probability evolution of packet incident on barrier with $m<E<V-m$.}
\label{fig:barrier_klein1}
\end{figure}

\begin{figure}[h!] 
\centering
\includegraphics[width=0.8 \textwidth]{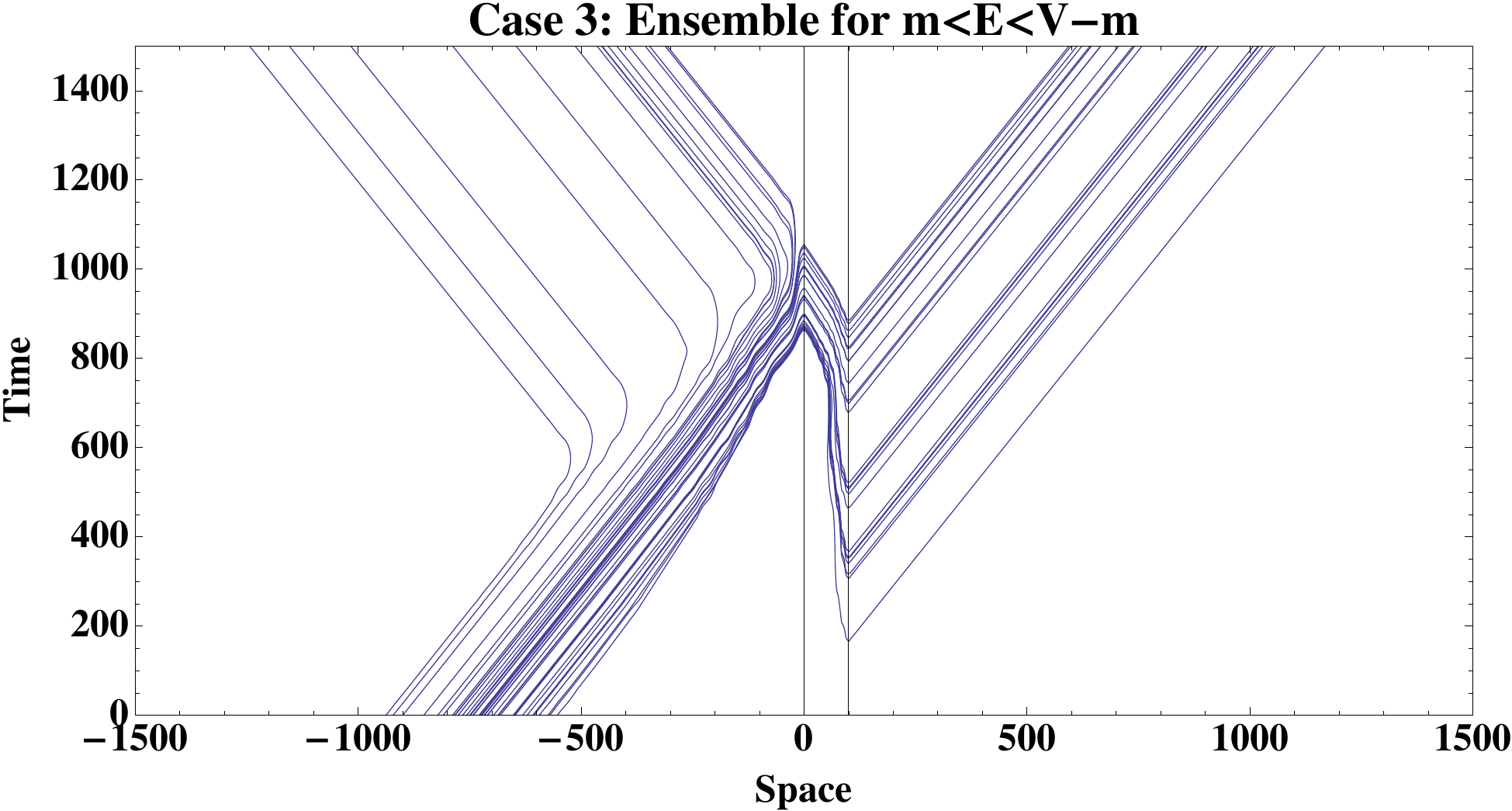}
\caption{(Case 3) Ensemble of trajectories for $m<E<V-m$ barrier with initial conditions at $t=0$ for random Gaussian distribution.}
\label{fig:barrier_klein2}
\end{figure}

Lastly we perform a simulation for Case 3: $m<E<V-m$ (Figure \ref{fig:barrier_klein1} \& \ref{fig:barrier_klein2}) taking $V=3$ and barrier width $L=100$. The Fabry-Perot pattern described in Section \ref{section:barrier} is visible as ripples in the probability density. While the packet inside the barrier appears to contradict the initial wavefunction on the $t=0$ time slice, it should be kept in mind that this is due to the finite size of the box used for the simulation. The packet that travels backwards in time in Region II diminishes in size as $t \rightarrow -\infty$ such that the initial wavefunction as $t  \rightarrow - \infty$ is a Gaussian packet in Region I.  For the purposes of numerical verification of probability conservation, we must add:
\begin{equation*}
P_B = \int_{0}^L  \Psi^\dagger(x,0) \Psi(x,0)  dx \quad \Rightarrow \quad P_R + P_T + P_B = P_A
\end{equation*} 
corresponding to conservation with respect to affine parameter $s$. By increasing the box size and moving the initial packet further from the barrier, $P_B \rightarrow 0$ and equality is approached. 

The particles which tunnel through the barrier appear to emerge in Region III \emph{before} the particle in Region I reached the potential boundary. According to the interpretation, this is due to the particle going backwards in time inside Region II. In the lab frame, this corresponds to a particle pair being produced at the right side of the barrier; the particle goes off to $x=+\infty$, and the anti-particle annihilates the incident particle on the left side of the boundary at a later time.  

As previously stated, particle trajectories cannot cross by uniqueness. It follows that a particle going backwards in time inside Region II cannot tunnel back into Region I and continue to $x=-\infty$. Instead it can be seen that incident particles are reflected well before reaching the barrier, and form distinct bands. By tracing the reflected trajectories of Figure \ref{fig:barrier_klein2} backwards towards the barrier, one sees that they agree with the spacetime points where particles would have tunneled back into Region I after multiple internal reflections in Region II (analogous to the interferometer). This is due to the fact that the wavefunction \emph{can} cross itself resulting in the Fabry-Perot pattern. Since the particles are \emph{guided} by the wave, this pattern is therefore visible in the trajectories.\cite{footnote3} Similar behavior occurs for the transmitted particles when tunneling out into Region III. The interpretation therefore produces the asymptotic behavior of the Fabry-Perot interferometer analogy.

\section{Conclusion}

We have shown that a consistent explanation can be given for a single relativistic particle process by combining the pilot wave and Feynman-Stueckelberg interpretations. The interpretation can be applied to scattering off a potential barrier as well as the Klein paradox. In this context, the original statement of the paradox then becomes a statement about pair production at the boundaries of high potentials. It would seem that the Klein step still has an interesting feature, even with this interpretation: the particle's initial condition is distributed with non-zero probability at a future time slice. This is due to the infinite extent of the potential $V>2mc^2$ yielding a negative energy solution at $x=+\infty$ - something that does not occur in the potential barrier. For both cases the wave traveling backwards in time interferes with itself; this causes constructive interference for the reflected packet in the Klein step, and a Fabry-Perot pattern for the Klein barrier. Future work can include the application of this interpretation to bosons where the Dirac sea picture breaks down, as well as a cohesive interpretation of a many particle process. Ascribing a continuous trajectory to the particle in this deterministic interpretation provides an intuitive perspective on particle creation/annihilation, and a resolution to the historic paradox.

\appendix*   

\section{Multiple Scattering Limit}

We show for the case of the Klein barrier that a packet going backwards in time inside Region II diminishes in size as $t \rightarrow - \infty$. This is done by showing the probability of an arbitrarily large number of internal reflections approaches zero. The probability of an incident packet being reflected without entering Region II is $(1-|D|^2)$. The probability of a transmitted packet not undergoing any internal reflection is $|D|^2 (1-|B|^2)$. For each higher order $n \in \{0,1, 2, ...\}$ of asymptotic reflection $\mathcal{R}(n)$ or transmission $\mathcal{T}(n)$ probability (corresponding to $+2$ internal reflections inside Region II), we multiply by $|D|^2 |B|^2$:
\begin{equation*}
\mathcal{R}(n) = \left( 1 - |D|^2 \right)  \left( |D|^2 |B|^2 \right)^n \quad \quad \& \quad \quad \mathcal{T}(n) = |D|^2 \left( 1 - |B|^2 \right) \left( |D|^2 |B|^2 \right)^n
\end{equation*}
\begin{equation*}
\Rightarrow \quad \sum_{n=0}^\infty \mathcal{R}(n) + \sum_{n=0}^\infty \mathcal{T}(n) = 1
\end{equation*}
where the number of internal reflections is $2n-1$ for asymptotically reflected packets, and $2n$ for asymptotically transmitted packets. We can solve for $|D|^2 |B|^2$ as:
\begin{equation*}
\left( |D|^2 |B|^2 \right) = \left( \frac{|T|^2}{4} \left[ 1 - \frac{1}{\kappa^2} \right]  \right)^2
\end{equation*}
For the probability of an arbitrary number of internal reflections to approach zero, we require:
\begin{equation*} 
\lim_{n \to \infty}   \left( |D|^2 |B|^2 \right)^n = 0 \quad \Rightarrow \quad \left( |D|^2 |B|^2 \right) < 1 \quad \Rightarrow \quad  \left| 1 - \frac{1}{\kappa^2} \right| < 1 \quad \Rightarrow \quad \kappa^2 \geq 1
\end{equation*}
since $|T|^2 \leq 1$. Using $V>2m$, $m<E<V-m$, and being careful with signs:
\begin{equation*}
\kappa \equiv \frac{k}{p} \frac{E+m}{E-V+m} \quad \quad \Rightarrow \quad \quad \kappa^2 =   \frac{E+m}{E-m}  \ \frac{ |E-V|+m }{ |E-V|-m }   \geq 1
\end{equation*}
We conclude the packet traveling backwards in time inside Region II decays as $t \rightarrow - \infty$.


\begin{acknowledgements}
The author is very grateful to Professor Allan S. Blaer for his guidance and invaluable discussions.
\end{acknowledgements}


\begin{thebibliography}{99}
\bibitem{footnote1} In the conventional approach we set $A=1$, even though it seems more natural in this interpretation to set $R=1$.
\bibitem{footnote2} This is because the wave moves backwards in time inside Region II by the Feynman-Stueckelberg interpretation.
\end{thebibliography}
\end{document}